%% file: sst_stego_article.tex
\documentclass[11pt,a4paper]{article}

\usepackage[margin=1in]{geometry}
\usepackage{amsmath,amssymb}
\usepackage{booktabs}
\usepackage{graphicx}
\usepackage{microtype}
\usepackage{pgfplots}
\usepackage{placeins}
\usepackage{float}
\usepackage{needspace}
\usepackage{hyperref}

\pgfplotsset{compat=1.18}

\title{Set Shaping Theory as a Complementary Payload-Shaping Layer for Steganography}
\author{Aida Koch\thanks{Contact: \texttt{aida.koch445@outlook.com}} \and Logan Lewis \and Lily Scott \and Agi Weber}
\date{\today}

\begin{document}

\maketitle

\begin{abstract}
This paper studies the use of Set Shaping Theory (SST) as a reversible
payload-shaping layer for least significant bit (LSB) image steganography. The
proposal is not intended to replace existing steganographic methods or compete
with them as a new embedding scheme. Instead, SST is positioned as a
complementary preprocessing stage that makes an existing embedding method easier
to apply with lower statistical disturbance. The transformation that applies set shaping theory by increasing the message length by k symbols is performed using the algorithm developed by Glen Tankersley which perform the transformation in an approximate but very fast way.
Although the embedded payload is lengthened from $N$ to $N+K$ bits, the selected
representation can reduce $D_{\mathrm{KL}}(P\Vert Q)$ and therefore make the
subsequent steganographic insertion less detectable under histogram-based
criteria. Across 1,800 controlled simulations on four synthetic cover-image
models, SST reduced $D_{\mathrm{KL}}(P\Vert Q)$ by an average of 25.16\%
relative to a fair $N+K$ LSB baseline, with a 95\% confidence interval of
$\pm$1.22\%. For $K=8$, the average reduction reached 42.81\%. Additional
robustness simulations with keyed random embedding paths confirmed the effect
across several distances: at $K=8$, SST reduced KL divergence by 42.44\%,
Jensen--Shannon divergence by 29.62\%, total variation by 12.41\%, and a
symmetric chi-square distance by 28.30\%. An additional image-based
matrix-embedding/STC-like simulation showed that SST also reduces the minimum
weighted insertion cost: relative to the unshaped $K=0$ reference, $K=8$
reduced the cost by 6.93\%.
\end{abstract}

\section{Introduction}

Modern steganography already contains many embedding strategies, ranging from
simple least significant bit (LSB) substitution to wet-paper coding, LSB
matching, syndrome-trellis coding, and distortion-minimizing adaptive schemes
\cite{wetpaper2005,mielikainen2006,filler2010,pevny2010hugo,filler2011,holub2014}.
The purpose of this paper is not to introduce a replacement for those methods.
Instead, it studies whether Set Shaping Theory (SST) can act as a
payload-shaping layer placed before an existing embedder. In this role, SST does
not decide where or how the cover is modified; it only chooses which reversible
representation of the same message should be given to the underlying
steganographic method.

LSB steganography is used here as a transparent testbed because its statistical
effect can be measured directly. It embeds a binary message by replacing the
least significant bits of selected pixels. The method is simple and
high-capacity, but its direct substitution mechanism perturbs the empirical
distribution of pixel intensities. Such perturbations can be measured by
comparing the original cover histogram $P$ with the stego histogram $Q$.

This work investigates whether SST can reduce that perturbation before the
message reaches the embedder. Rather than embedding the message in its original
form, the encoder constructs multiple reversible representations of the same
message. It then passes each candidate through the same embedding rule,
evaluates the induced divergence, and keeps the candidate whose histogram is
closest to the original image. The receiver uses the additional $K$ index bits
to invert the selected transformation and recover the original message.

The apparent trade-off is unusual: the payload becomes longer, from $N$ to
$N+K$, but the distributional damage produced by the existing embedder can
decrease. Therefore SST should be interpreted as a compatibility layer that can
make established steganographic techniques easier to apply under a statistical
distortion constraint, not as a competing steganographic system.

This article participates to Set Shaping Theory simulator project
available at \url{https://sst-simulator.github.io/Set-Shaping-Theory-Simulator/}.
Its steganography section provides an interactive implementation of the same
payload-shaping idea studied in this article, allowing the reader to vary the
message, the shaping parameter, and the embedding comparison in a direct visual
environment.

\section{Set Shaping Theory}

Classical information theory and source coding usually begin from the problem of
representing a source sequence with minimum expected length or minimum redundancy
\cite{shannon1948, huffman1952, witten1987}. In this framework, the central object is
the source model: if the probability law is known, one designs a code adapted to that
law; if the law is not known, universal methods attempt to approach the same limit
without assuming complete prior knowledge \cite{ziv1977, ziv1978}. Model-selection
principles such as minimum description length similarly connect representation quality
to the regularity that can be extracted from the data and from the model used to
describe it \cite{rissanen1978}. In all these approaches, the dominant question is
therefore dimensional: how many bits or symbols are needed to represent the data?

Set Shaping Theory (SST) starts from a different but compatible question. Instead of
changing only the code used to describe a fixed sequence, SST studies transformations
that change the combinatorial set in which the sequence is represented. In its original
form, SST was introduced as the study of bijective maps that transform a set of strings
of length \(N\) into a subset of strings of greater length \(N+K\), while preserving the
number of admissible messages \cite{kozlov2021}. If the alphabet is denoted by
\(\mathcal{A}\), with \(|\mathcal{A}|=A\), the basic object can be written as
\begin{equation}
	f:\mathcal{A}^{N} \longrightarrow \mathcal{Y}_{N+K},
	\qquad
	\mathcal{Y}_{N+K} \subset \mathcal{A}^{N+K},
	\qquad
	|\mathcal{Y}_{N+K}| = |\mathcal{A}^{N}| = A^{N}.
\end{equation}
Thus, SST does not simply append arbitrary redundant symbols. It selects a structured
subset \(\mathcal{Y}_{N+K}\) of the larger ambient space and creates a one-to-one
correspondence between the original messages and the admissible longer
representations.

In the compression-oriented formulation, the selected subset is chosen so that the
average information content, or equivalently the empirical entropy--length product,
is reduced. If \(H_0(x)\) denotes the zero-order empirical entropy of a sequence \(x\),
one of the characteristic SST objectives is
\begin{equation}
	\left\langle (N+K) H_0(f(s)) \right\rangle
	<
	\left\langle N H_0(s) \right\rangle,
	\qquad s \in \mathcal{A}^{N}.
\end{equation}
This inequality expresses the central intuition of SST: a longer representation can
still be more structured, and therefore more favorable for a later coding or processing
operation, than the original shorter representation. The relevant quantity is not
length alone, but the interaction between length and structure.

The same idea was later explored in several related directions. In locally testable
codes, SST was used to introduce dependencies before the final coding operation, so
that inadmissible decoded sequences can reveal errors through the fact that they do not
belong to the shaped image of the transformation \cite{kozlov2022}. In Huffman coding,
SST was studied as a practical preprocessing operation, with particular attention to
the difficulty of implementing the full correspondence table, whose size grows as
\(A^N\) \cite{schmidt2022}. Subsequent work emphasized the role of the shaping order
\(K\), which controls the size of the candidate representation family, and investigated
the relation between SST and empirical entropy coding limits \cite{biereagu2023}. More recent work on redundancy-free testable codes further develops the
idea that sequence extension can create structural constraints without necessarily
behaving like conventional explicit redundancy \cite{koch2025}.

For the present paper, the most important point is that SST should not be interpreted
only as a compression technique. Compression is one application in which the objective
function is related to \(H_0\), coding length, or empirical information content. More
generally, SST can be viewed as a data-conditioning principle: given many reversible
representations of the same underlying message, choose the representation that is most
compatible with the downstream operation. In this broader interpretation, the
objective function need not be entropy. It can be any measurable cost induced by the
system that will process the shaped data.

This distinction is essential in steganography. The goal of the present article is not
to reduce the number of embedded bits and not to replace the steganographic embedder.
The goal is to choose, among several reversible payload representations, the one that
causes the smallest statistical disturbance when passed through the same embedding
method. Therefore, the shaped set is evaluated with respect to the cover stego
relationship rather than with respect to compression alone.

In this sense, SST acts as a complementary payload-shaping layer. It does not decide
where the cover should be modified, and it does not define a new extraction rule. Those
tasks remain the responsibility of the underlying steganographic method. SST only
changes the representation of the payload before embedding, so that the same method
can operate on an input whose structure is more favorable under the chosen distortion
criterion. This makes SST cooperative rather than competitive with existing
steganographic systems: it is an upstream reversible preprocessing stage designed to
reduce the statistical cost paid by a downstream embedder.

\section{Method}

Let $x=(x_1,\ldots,x_M)$ be a grayscale cover image represented as pixel values
in $\{0,\ldots,255\}$, and let $s\in\{0,1\}^N$ be the secret message. Let
$\mathcal{E}$ denote an existing steganographic embedder. In the experiments,
$\mathcal{E}$ is the direct LSB rule, which produces a stego image by replacing
the least significant bit of the first $N$ pixels:
\begin{equation}
    y_i = \operatorname{LSBSet}(x_i,s_i), \qquad i=1,\ldots,N.
\end{equation}
The SST component is inserted before $\mathcal{E}$ and leaves the embedding
rule itself unchanged.
The Set Shaping Theory will be applied using the developed method by Glen Tankersley\cite{glen2025} . For SST-assisted embedding, a shaping overhead $K$ defines a family of
$H=2^K$ reversible binary transformations. For each index $h\in\{0,\ldots,H-1\}$,
a deterministic binary mask $r_h\in\{0,1\}^N$ is generated and the transformed
message is
\begin{equation}
    s^{(h)} = s \oplus r_h.
\end{equation}
The embedded payload is the concatenation
\begin{equation}
    z^{(h)} = \operatorname{bin}_K(h) \Vert s^{(h)},
\end{equation}
where $\operatorname{bin}_K(h)$ is the $K$-bit binary representation of the
chosen index. This payload is then passed to the same underlying embedder
$\mathcal{E}$. The decoder first extracts the payload by the ordinary decoding
procedure of that embedder, reads the index, regenerates the corresponding mask,
and recovers
\begin{equation}
    s = s^{(h)} \oplus r_h.
\end{equation}

For each candidate $h$, the encoder computes the empirical histogram $Q_h$ of
the corresponding stego image and minimizes the Kullback-Leibler divergence
\begin{equation}
    D_{\mathrm{KL}}(P\Vert Q_h)
    =
    \sum_{v=0}^{255} P(v)\log_2\frac{P(v)}{Q_h(v)}.
\end{equation}
Small additive smoothing is used only to avoid undefined logarithms when a bin
has zero empirical mass.

This formulation separates payload shaping from steganographic embedding. Any
method that can embed and recover a binary payload could, in principle, receive
an SST-shaped payload. LSB is therefore not the proposed final system; it is the
controlled carrier used to quantify the shaping effect.

\section{Experimental Design}

The simulation campaign uses $100\times100$ grayscale images, binary messages
of length $N\in\{1000,2500,4000\}$, and shaping overhead
$K\in\{0,2,4,6,8\}$. Four synthetic cover-image models are tested:
uniform random images, smoothed random images, noisy gradients, and bimodal
mixtures. Each configuration is repeated 30 times, yielding 1,800 runs. The
goal is to isolate whether the shaping layer reduces the distortion caused by a
fixed embedder, not to claim that LSB itself is the strongest steganographic
choice.

The primary baseline is a fair LSB insertion of $N+K$ bits: the first $K$ bits
are a zero index and the remaining $N$ bits are the unshaped message. This is
the correct comparison because the SST-shaped payload also has length $N+K$ and
is embedded by the same LSB rule. The main reported metric is the relative gain
\begin{equation}
    G =
    \frac{D_{\mathrm{KL}}(P\Vert Q_{\mathrm{base}}) -
          D_{\mathrm{KL}}(P\Vert Q_{\mathrm{SST}})}
         {D_{\mathrm{KL}}(P\Vert Q_{\mathrm{base}})}.
\end{equation}
Positive values indicate that SST produced a stego histogram closer to the
original cover histogram while using the same underlying embedding mechanism.
This is aligned with information-theoretic steganography, where relative
entropy can bound the distinguishability between cover and stego distributions
\cite{cachin2004}.

Two additional simulation blocks were then added to test robustness. First, the
embedding path was randomized by a deterministic key rather than fixed to the
first $N+K$ pixels. The same fair $N+K$ baseline was used, and the selected SST
candidate was still chosen by minimizing KL divergence. The resulting stego
images were then evaluated with KL divergence, Jensen--Shannon divergence,
total variation, symmetric chi-square distance
$\sum_v (P(v)-Q(v))^2/(P(v)+Q(v))$, and the L1 distance between horizontal
pixel-pair co-occurrence matrices. Second, a smaller timing study measured the
cost of exhaustive candidate search for $K\in\{0,4,8,10,12\}$.

A third simulation tested the same complementarity claim with a
matrix-embedding/STC-like syndrome encoder applied to generated grayscale
images. Pixel positions were selected from the image by a keyed random path,
their LSBs formed the binary cover sequence, and local image texture determined
the weighted cost of changing each selected pixel. The selected positions were
then divided into syndrome blocks, and dynamic programming was used to compute
the minimum weighted flip cost required to realize a target syndrome. SST tested
the equivalent payloads $\operatorname{bin}_K(h)\Vert(s\oplus r_h)$ and selected
the one with minimum syndrome-embedding cost. The case $K=0$ has no shaping
choice and is used as the reference. The decoded message is therefore unchanged;
only the representation given to the matrix embedder changes.

\clearpage
\section{Results}

\Needspace{0.30\textheight}
The central effect of the shaping overhead is reported in Table~\ref{tab:gain-k}.
\input{sst_stego_article_results/table_gain_k.tex}
With $K=0$, SST has no additional choice and matches the fair baseline. Once
$K>0$, the average gain becomes positive and increases monotonically with the
size of the candidate set. At $K=8$, the encoder tests 256 reversible
representations and obtains a 42.81\% mean reduction in
$D_{\mathrm{KL}}(P\Vert Q)$.

\Needspace{0.50\textheight}
The same monotone trend is shown visually in Figure~\ref{fig:gain-by-k}.
\input{sst_stego_article_results/figure_gain_by_k.tex}
The error bars indicate that the observed gain is not limited to a single
configuration, and the increasing curve supports the interpretation that the
additional SST choices are doing useful work.

\Needspace{0.50\textheight}
Figure~\ref{fig:kl-comparison} compares the absolute KL values of the fair
baseline and the best SST-shaped payload.
\input{sst_stego_article_results/figure_kl_comparison.tex}
The fair baseline remains roughly stable because it embeds the same length
$N+K$ without candidate selection, while the best SST candidate steadily lowers
the final histogram divergence as $K$ increases.

\Needspace{0.30\textheight}
The dependence on the cover-image model is summarized in
Table~\ref{tab:gain-cover}.
\input{sst_stego_article_results/table_gain_cover.tex}
Smooth and bimodal images show larger average gains than uniform and gradient
images. This is plausible because structured histograms provide more room for
candidate selection to avoid visibly unbalanced bin changes.

\Needspace{0.50\textheight}
The same cover-model dependence is displayed in Figure~\ref{fig:gain-by-cover}.
\input{sst_stego_article_results/figure_gain_by_cover.tex}
The figure is useful as a practical reading of the table: SST is most effective
when the cover distribution has structure that the candidate search can exploit.

\Needspace{0.30\textheight}
Table~\ref{tab:gain-n} reports the effect across the tested message lengths.
\input{sst_stego_article_results/table_gain_n.tex}
The gain remains close to 25\% for all three values of $N$, which suggests that
the result is not caused by a single payload size. These results should be read
as evidence that SST can reduce the statistical burden placed on an existing
steganographic method.

\Needspace{0.35\textheight}
The robustness simulations with keyed random embedding paths are reported in
Table~\ref{tab:robust-metrics}.
\input{sst_stego_robustness_results/table_robust_metrics.tex}
The main KL result is preserved when the embedding locations are selected by a
key rather than by the first pixels of the image. At $K=8$, the KL reduction is
42.44\%, very close to the sequential-path result. The same shaped payload also
improves Jensen--Shannon divergence, total variation, and symmetric chi-square
distance, although the improvement is smaller for co-occurrence statistics.
This is expected: the SST objective in the present experiments is
histogram-based, so spatial pair statistics are affected only indirectly.

\Needspace{0.30\textheight}
The distribution of the selected shaping index is shown in
Table~\ref{tab:h-distribution}.
\input{sst_stego_robustness_results/table_h_distribution.tex}
The selected index is not concentrated on a single fixed value. For $K=8$, the
mean normalized index is 0.517 and the largest selected-index bucket accounts
for only 2.08\% of runs. Thus the gain comes from genuine candidate selection
rather than from a particular mask dominating all covers.

\Needspace{0.30\textheight}
The computational cost of larger shaping orders is given in
Table~\ref{tab:cost-k}.
\input{sst_stego_robustness_results/table_cost_k.tex}
The table shows the expected exponential cost in $K$, but also that the
per-candidate time remains roughly constant. In the tested Node.js
implementation, exhaustive search with $K=12$ requires about 165 ms for
$N=1000$ on the selected structured cover models.

\Needspace{0.35\textheight}
The image-based matrix-embedding/STC-like experiment is summarized in
Table~\ref{tab:stc-sst-cost}.
\input{sst_stc_matrix_results/matrix_embedding_table.tex}
Here the objective is no longer histogram divergence but the minimum weighted
insertion cost needed to obtain the same recoverable message. The result is
consistent with the main thesis: increasing $K$ gives the encoder more
equivalent message representations, and the minimum achievable embedding cost
decreases as the candidate set grows. Relative to the $K=0$ reference, $K=8$
reduces the minimum insertion cost
by 6.93\% over 960 image-based runs.

\FloatBarrier

\section{Discussion}

The experiments support the central hypothesis of this work: Set Shaping Theory can
serve as an effective complementary payload-shaping layer for steganography. The
important point is that SST is not proposed as a replacement for existing
steganographic methods. It does not compete with LSB embedding, matrix embedding,
syndrome-trellis coding, adaptive embedding, or other advanced schemes. Instead, it
acts before the embedding stage, selecting a more favorable reversible representation
of the same message. In this sense, SST improves the input given to the embedder rather
than changing the embedder itself.

This distinction is crucial. The results show that even when the payload length is
increased from \(N\) to \(N+K\), the final statistical disturbance can decrease
significantly. In the LSB experiments, the SST-shaped payload produced a substantial
reduction in \(D_{\mathrm{KL}}(P\|Q)\) relative to the fair \(N+K\) baseline. The gain
increased systematically with \(K\), reaching more than \(40\%\) reduction in KL
divergence for \(K=8\). This means that the additional representation choices created
by SST are not merely redundant overhead; they provide useful degrees of freedom that
allow the encoder to select a payload representation more compatible with the cover
image.

The robustness tests reinforce this interpretation. When the embedding positions were
selected by keyed random paths, the same positive trend remained visible across several
statistical distances, including KL divergence, Jensen--Shannon divergence, total
variation, and symmetric chi-square distance. This indicates that the effect is not
limited to a single sequential embedding path or to a single metric. The shaped payload
is generally less disruptive under several histogram-based criteria, which is precisely
the desired behavior for a preprocessing layer intended to support steganographic
embedding.

The matrix-embedding/STC-like experiment is particularly important because it moves
the interpretation beyond simple LSB substitution. In that setting, the objective is
not only to reduce a histogram divergence after embedding, but to reduce the minimum
weighted insertion cost required by an existing syndrome-based embedder. The observed
decrease in insertion cost shows that SST can be useful even when the downstream method
already performs an optimization. This supports the view that SST is not limited to
basic steganographic schemes; rather, it can provide an additional optimization layer
on top of more advanced methods.

Although LSB steganography is used in this article as a transparent and easily
measurable testbed, the principle is more general. Any steganographic method that
embeds a binary payload can, in principle, receive multiple reversible representations
of the same message and choose the one that minimizes a relevant distortion function.
For LSB, this function can be a histogram divergence. For adaptive steganography, it
could be a spatial distortion cost. For syndrome-trellis coding, it could be the
weighted embedding cost. For more advanced systems, it could be a detector-based or
model-based security score. The common structure is the same: SST increases the number
of admissible payload representations and allows the encoder to select the one that is
most compatible with the cover and with the embedding rule.

Therefore, the main contribution of this work is not the proposal of a new
steganographic algorithm in competition with existing ones, but the introduction of a
general reversible shaping layer that can cooperate with them. The results suggest that
important gains can be obtained simply by improving the representation of the payload
before embedding. This makes SST a promising tool for steganographic systems in which
the message can be preprocessed and where the encoder is allowed to choose among
equivalent representations in order to reduce statistical detectability or embedding
cost.

\section{Conclusion}

This paper has presented Set Shaping Theory as a complementary payload-shaping layer
for steganography. SST is not proposed as a replacement for existing embedding
methods, but as a reversible preprocessing stage that improves the representation of
the payload before embedding.

The results show that this approach can produce significant gains. In the LSB
experiments, SST reduced the KL divergence between the cover and stego histograms,
with the improvement exceeding \(40\%\) for \(K=8\). The same positive trend appeared
with keyed random embedding paths and with other statistical distances, confirming
that the effect is not limited to one specific insertion path.

The matrix-embedding/STC-like experiment further shows that SST can also reduce the
minimum weighted insertion cost required by a more advanced embedding mechanism. This
supports the main interpretation of the method: SST can cooperate with existing
steganographic systems by giving them a more favorable payload representation.

Although LSB is used here as the main testbed, the principle is more general. Any
steganographic method that receives a binary payload can potentially benefit from a
reversible shaping layer that selects the most compatible representation of the same
message. For this reason, SST should be viewed as a general payload-conditioning
method, suitable for both simple and advanced steganographic schemes.

\appendix

\section{Online Set Shaping Theory Simulator}

This article is part of the Set Shaping Theory simulator project, available
online at
\url{https://sst-simulator.github.io/Set-Shaping-Theory-Simulator/}. The project is
not only a demonstration tool, but also a research site intended to collect
examples, use cases, and problem instances in which the structure of the data is more
important than its raw length. 

The steganography section of the simulator focuses on the same complementary role
studied in this paper. The user provides or generates a binary message and selects a
shaping parameter \(K\). The simulator then applies Set Shaping Theory to the message producing a message with k more inputs but with a much smaller Kullback-Leibler divergence. In this way, the simulator
shows that SST does not hide the message by itself and does not replace the
underlying steganographic method. Instead, it changes the representation given to
that method so that the subsequent embedding step can be easier, less statistically
disturbing, or less costly.

The online tool therefore provides a practical way to reproduce the central idea of
the article: the relevant comparison is always against the unshaped \(K=0\)
reference, while larger values of \(K\) give the encoder more equivalent payload
representations from which to choose. The simulator is especially useful for
visualizing why a longer shaped payload can still reduce divergence or insertion
cost when the chosen representation is better matched to the cover data and to the
embedding rule.

\end{document}

%% file: sst_stego_article_results/table_gain_k.tex
% Extracted from article_table_snippets.tex for local Results layout.
\begin{table}[H]
\centering
\caption{Mean reduction in $D_{\mathrm{KL}}$ relative to the fair $N+K$ LSB baseline as a function of shaping overhead.}
\label{tab:gain-k}
\begin{tabular}{lrrrr}
\toprule
$K$ & Runs & Mean gain & 95\% CI & Success \\
\midrule
0 & 360 & 0.00\% & $\pm$ 0.00\% & 0.00\% \\
2 & 360 & 17.69\% & $\pm$ 3.08\% & 81.11\% \\
4 & 360 & 28.68\% & $\pm$ 2.45\% & 91.94\% \\
6 & 360 & 36.61\% & $\pm$ 2.25\% & 98.06\% \\
8 & 360 & 42.81\% & $\pm$ 2.15\% & 100.00\% \\
\bottomrule
\end{tabular}
\end{table}

%% file: sst_stego_article_results/figure_gain_by_k.tex
% Extracted from article_plot_snippets.tex for local Results layout.
\begin{figure}[H]
\centering
\begin{tikzpicture}
\begin{axis}[
  width=0.78\linewidth,
  height=0.43\linewidth,
  xlabel={$K$},
  ylabel={Gain vs. fair LSB (\%)},
  grid=both,
  ymin=0
]
\addplot+[mark=o, thick, error bars/.cd, y dir=both, y explicit] coordinates {
(0,0.0000) +- (0,0.0000)
(2,17.6926) +- (0,3.0793)
(4,28.6793) +- (0,2.4487)
(6,36.6108) +- (0,2.2543)
(8,42.8054) +- (0,2.1487)
};
\end{axis}
\end{tikzpicture}
\caption{Mean KL-divergence reduction obtained by SST as the number of shaping index bits increases. Error bars show approximate 95\% confidence intervals.}
\label{fig:gain-by-k}
\end{figure}
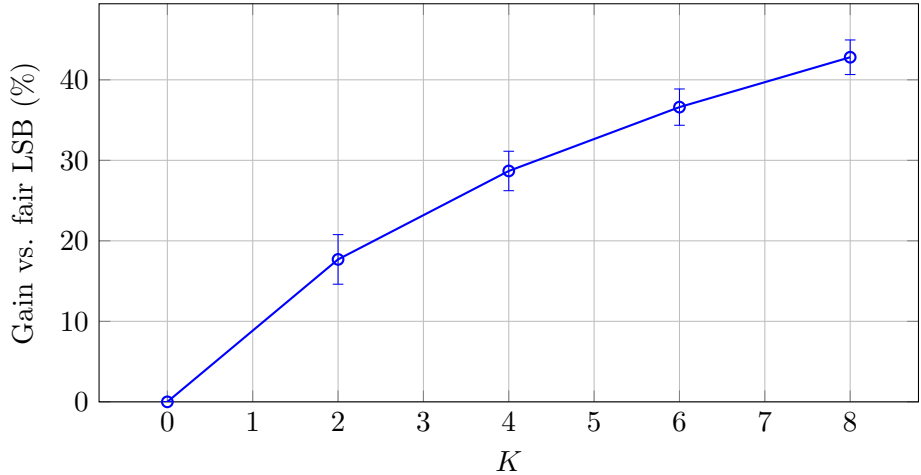

%% file: sst_stego_article_results/figure_kl_comparison.tex
% Extracted from article_plot_snippets.tex for local Results layout.
\begin{figure}[H]
\centering
\begin{tikzpicture}
\begin{axis}[
  width=0.78\linewidth,
  height=0.43\linewidth,
  xlabel={$K$},
  ylabel={Mean $D_{\mathrm{KL}}(P\Vert Q)$},
  grid=both,
  legend pos=north east
]
\addplot+[mark=o, thick] coordinates {
(0,8.954637e-3)
(2,9.007500e-3)
(4,8.641641e-3)
(6,8.791135e-3)
(8,8.689698e-3)
};
\addlegendentry{Fair LSB baseline}
\addplot+[mark=square, thick] coordinates {
(0,8.954637e-3)
(2,6.666066e-3)
(4,5.520419e-3)
(6,4.846998e-3)
(8,4.340269e-3)
};
\addlegendentry{Best SST}
\end{axis}
\end{tikzpicture}
\caption{Average KL divergence for the fair $N+K$ LSB baseline and for the best SST-shaped payload.}
\label{fig:kl-comparison}
\end{figure}
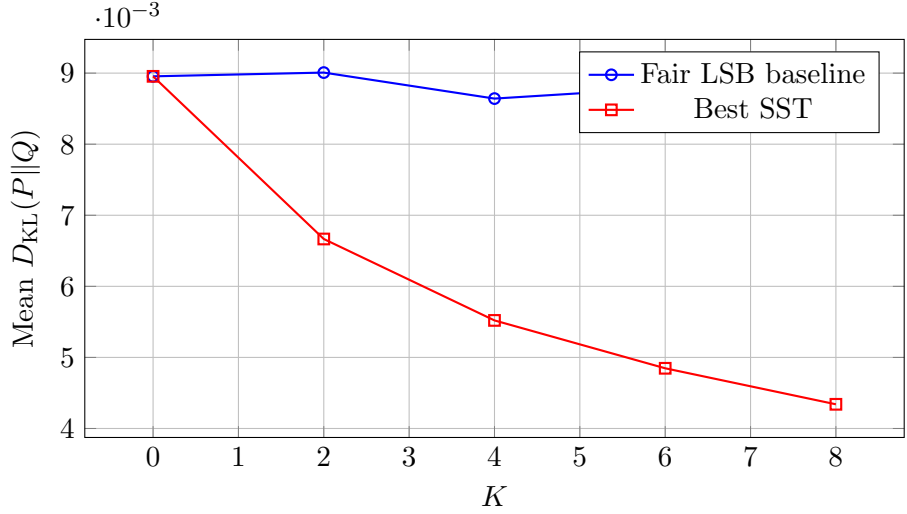

%% file: sst_stego_article_results/table_gain_cover.tex
% Extracted from article_table_snippets.tex for local Results layout.
\begin{table}[H]
\centering
\caption{Mean reduction in $D_{\mathrm{KL}}$ by synthetic cover-image model.}
\label{tab:gain-cover}
\begin{tabular}{lrrrr}
\toprule
Cover model & Runs & Mean gain & 95\% CI & Success \\
\midrule
uniform & 450 & 15.87\% & $\pm$ 1.24\% & 74.22\% \\
smooth & 450 & 36.73\% & $\pm$ 3.04\% & 74.67\% \\
gradient & 450 & 16.78\% & $\pm$ 1.45\% & 73.78\% \\
bimodal & 450 & 31.25\% & $\pm$ 2.87\% & 74.22\% \\
\bottomrule
\end{tabular}
\end{table}

%% file: sst_stego_article_results/figure_gain_by_cover.tex
% Extracted from article_plot_snippets.tex for local Results layout.
\begin{figure}[H]
\centering
\begin{tikzpicture}
\begin{axis}[
  ybar,
  width=0.78\linewidth,
  height=0.43\linewidth,
  symbolic x coords={uniform,smooth,gradient,bimodal},
  xtick=data,
  ylabel={Gain vs. fair LSB (\%)},
  xlabel={Cover model},
  grid=both,
  ymin=0
]
\addplot coordinates {
(uniform,15.8704)
(smooth,36.7314)
(gradient,16.7775)
(bimodal,31.2512)
};
\end{axis}
\end{tikzpicture}
\caption{Mean SST gain by cover-image model.}
\label{fig:gain-by-cover}
\end{figure}
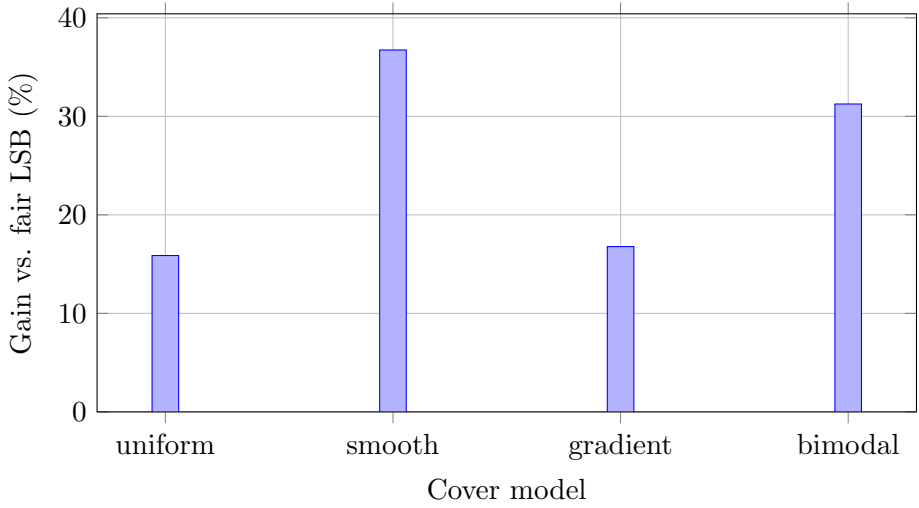

%% file: sst_stego_article_results/table_gain_n.tex
% Extracted from article_table_snippets.tex for local Results layout.
\begin{table}[H]
\centering
\caption{Mean reduction in $D_{\mathrm{KL}}$ by embedded message length.}
\label{tab:gain-n}
\begin{tabular}{lrrrr}
\toprule
$N$ & Runs & Mean gain & 95\% CI & Success \\
\midrule
1000 & 600 & 24.87\% & $\pm$ 2.40\% & 74.17\% \\
2500 & 600 & 25.98\% & $\pm$ 1.99\% & 74.33\% \\
4000 & 600 & 24.62\% & $\pm$ 1.94\% & 74.17\% \\
\bottomrule
\end{tabular}
\end{table}

%% file: sst_stego_robustness_results/table_robust_metrics.tex
% Extracted from robustness_table_snippets.tex for local Results layout.
\begin{table}[H]
\centering
\caption{Robustness check with keyed random embedding paths. Mean reduction relative to the fair $N+K$ baseline is reported for several statistical distances.}
\label{tab:robust-metrics}
\begin{tabular}{lrrrrr}
\toprule
$K$ & $D_{\mathrm{KL}}$ & JS & TV & $\chi^2$ & Cooc. L1 \\
\midrule
0 & 0.00\% & 0.00\% & 0.00\% & 0.00\% & 0.00\% \\
2 & 16.72\% & 9.47\% & 2.53\% & 8.77\% & 0.26\% \\
4 & 28.72\% & 17.15\% & 6.63\% & 16.03\% & 0.92\% \\
6 & 37.06\% & 23.80\% & 9.05\% & 22.50\% & 1.31\% \\
8 & 42.44\% & 29.62\% & 12.41\% & 28.30\% & 1.49\% \\
\bottomrule
\end{tabular}
\end{table}

%% file: sst_stego_robustness_results/table_h_distribution.tex
% Extracted from h_distribution_table.tex for local Results layout.
\begin{table}[H]
\centering
\caption{Distribution of selected shaping index $h$. Values close to $0.5$ for normalized $h$ and small maximum-bucket shares indicate that gains are not caused by a single fixed index.}
\label{tab:h-distribution}
\begin{tabular}{rrrr}
\toprule
$K$ & Runs & Mean normalized $h$ & Largest bucket share \\
\midrule
2 & 240 & 0.500 & 28.33\% \\
4 & 240 & 0.463 & 9.17\% \\
6 & 240 & 0.518 & 3.33\% \\
8 & 240 & 0.517 & 2.08\% \\
\bottomrule
\end{tabular}
\end{table}

%% file: sst_stego_robustness_results/table_cost_k.tex
% Extracted from robustness_table_snippets.tex for local Results layout.
\begin{table}[H]
\centering
\caption{Computational cost for larger shaping orders. The timing measures exhaustive candidate search with keyed random paths, $N=1000$, and two structured cover models.}
\label{tab:cost-k}
\begin{tabular}{rrrr}
\toprule
$K$ & Configurations & Mean search time (ms) & Mean time/candidate ($\mu$s) \\
\midrule
0 & 1 & 0.08 & 77.17 \\
4 & 16 & 0.64 & 40.23 \\
8 & 256 & 9.72 & 37.97 \\
10 & 1024 & 40.04 & 39.10 \\
12 & 4096 & 165.31 & 40.36 \\
\bottomrule
\end{tabular}
\end{table}

%% file: sst_stc_matrix_results/matrix_embedding_table.tex
% Auto-generated by sst_matrix_embedding_simulation.js
\begin{table}[H]
\centering
\caption{Matrix-embedding/STC-like minimum insertion cost on generated grayscale images after SST payload shaping. The reference is the unshaped case $K=0$; larger $K$ values give the encoder more equivalent payload representations to choose from.}
\label{tab:stc-sst-cost}
\begin{tabular}{rrrrr}
\toprule
$K$ & Configurations & Runs & Minimum cost & Reduction vs. $K=0$ \\
\midrule
0 & 1 & 960 & 160.092 & 0.00\% \\
2 & 4 & 960 & 157.473 & 1.64\% \\
4 & 16 & 960 & 154.018 & 3.79\% \\
6 & 64 & 960 & 151.312 & 5.48\% \\
8 & 256 & 960 & 148.996 & 6.93\% \\
\bottomrule
\end{tabular}
\end{table}